\documentclass[prd,nofootinbib,superscriptaddress,twocolumn]{revtex4-2}

\usepackage{amsfonts,amssymb,amsthm,bbm,amsmath}
\usepackage{hyperref}

\usepackage{color}

\usepackage{tikz}
\usetikzlibrary{calc}
\usetikzlibrary{decorations.pathmorphing}
\usepackage{subcaption}
\newcommand{\C}{{\mathbb C}}

\newcommand{\R}{{\mathbb R}}

\newcommand{\cK}{{\mathcal K}}
\newcommand{\cL}{{\mathcal L}}
\newcommand{\cH}{{\mathcal H}}

\newcommand{\cS}{{\mathcal S}}

\newcommand{\SL}{\mathrm{SL}}

\newcommand{\be}{\begin{equation}}
\newcommand{\ee}{\end{equation}}
\newcommand{\beq}{\begin{eqnarray}}
\newcommand{\eeq}{\end{eqnarray}}
\newcommand{\bes}{\begin{eqnarray}}
\newcommand{\ees}{\end{eqnarray}}

\renewcommand{\sl}{{\mathfrak{sl}}}

\newcommand{\la}{\langle}
\newcommand{\ra}{\rangle}

\newcommand{\f}{\frac}

\def\nn{\nonumber}
\def\pp{\partial}

\def\rd{\mathrm{d}}

\def\eps{\epsilon}

\def\om{\omega}
\def\Om{\Omega}

\def\tt{\tilde{t}}
\def\tq{\tilde{q}}
\def\tom{\tilde{\om}}
\def\bpsi{\bar{\psi}}

\begin{document}

\title{Quantum Uncertainty as an Intrinsic Clock}

\author{{\bf Etera R. Livine}}\email{etera.livine@ens-lyon.fr}
\affiliation{Laboratoire de Physique, ENS Lyon, CNRS-UMR 5672, 46 all\'ee d'Italie, Lyon 69007, France}
\affiliation{Perimeter Institute for Theoretical Physics,  Waterloo, Ontario N2L 2Y5, Canada}

\date{\today}

\begin{abstract}

In quantum mechanics, a classical particle is raised to a wave-function, thereby acquiring many more degrees of freedom. For instance, in the semi-classical regime, while the position and momentum expectation values follow the classical trajectory, the uncertainty of a wave-packet can evolve and beat independently.  We use this insight to revisit the dynamics of a 1d particle in a time-dependent harmonic well. One can solve it by considering time reparameterizations and the Virasoro group action to map the system to the harmonic oscillator with constant frequency. We prove that identifying such a simplifying time variable is naturally solved by quantizing the system and looking at the evolution of the width of a Gaussian wave-packet.
We further show that the Ermakov-Lewis invariant for the classical evolution in a time-dependent harmonic potential is actually the quantum uncertainty of a Gaussian wave-packet. This naturally extends the classical Ermakov-Lewis invariant to a constant of motion for quantum systems following Schr\"odinger equation.
We conclude with a discussion of potential applications to quantum gravity and quantum cosmology.

\end{abstract}

\maketitle




Systems evolving in time-dependent potentials are ubiquitous in physics. They appear in diverse settings, from Floquet oscillators in condensed matter to matter field propagations in an expanding universe in cosmology. Their study and applications are essential both from a theoretical and an experimental standpoint.
This short paper is dedicated to revisiting the simplest of such systems, that is the one-dimension particle evolving in a time-dependent harmonic potential, with the aim of underlining the role of time reparameterizations and conserved charges in both the classical and the quantum theories.

Indeed, having time-dependent potential and Hamiltonian naturally opens the door to playing with the choice of time coordinate and to the possibility of choosing a more suitable clock that the original time to describe the evolution of the system. Mathematically, it amounts to  considering possible changes of the time variable in order to simplify and integrate the equation of motion.
Our starting point is the remark that time reparameterizations for a harmonic potential in classical mechanics lead to the Virasoro group action on the potential with the Schwarzian term appearing in the action.
This allows to choose a specific new clock with respect to which the explicit time dependence disappears. The system is then mapped onto a standard harmonic oscillator with constant frequency. The simplifying new time variable must satisfy a non-linear differential equation. Assuming that such a simplifying clock has been identified, this construction then allows to derive conserved charges, which are known as the Ermakov-Lewis constants of motion for the 1d particle in a time-dependent quadratic potential \cite{Lewis,Leach,LewisLeach,Schuch:2008tha,Gallegos2018,Padmanabhan2018}.
In this context, we point out that these invariants are the squared Wronskians evaluated on the classical trajectory and show that they generate, by Noether theorem, the classical $\SL(2,\R)$ symmetry under M\"obius reparameterization of the time coordinate.

Things get even more interesting at the quantum level, with the interlaced evolution of the classical system and its quantum fluctuations.
First, we review the exact dynamics of a Gaussian wave-packet in a time-dependent harmonic potential, which describes the the semi-classical regime of the quantum theory. We can then show that the evolution of the width (in position) of the Gaussian state satisfies the required classical non-linear differential equation and thus provides the wanted simplifying clock to integrate the classical equation of motion.
This is a surprising example of a classical problem finding a natural quantum solution. Physically, this means that one focuses on the evolution of the quantum fluctuations of the wave-packet and uses them as a clock to describe the evolution of the position and momentum expectation values along the classical trajectory. In some sense, the wave-packet is like a beating heart, and the classical system in terms of that ``quantum'' beat evolves as a harmonic oscillator with constant frequency.
Pushing further this classical-quantum interplay, we recall that the quantum uncertainty, defined \`a la Heisenberg as the product of the quantum uncertainties in position and momentum, is a conserved charge of the quantum system as a direct consequence of the Schr\"odinger equation satisfied by  the wave-function. And we show that the quantum uncertainty of a Gaussian wave-packet is exactly the Ermakov-Lewis invariant of the classical trajectory.
This is a surprising example of a quantum observable playing the role of the generator of the symmetry of the classical system.

We conclude this letter with a short discussion of possible applications of such classical-quantum interplay to quantum cosmology and quantum gravity.

\section{Time-dependent Potential}

Let us look into the classical particle evolving in a time-dependent harmonic potential. This system is driven by the following action:
\be
s_{\om}[t,q(t)]
=\int \rd t \left[
\f12m\dot{q}^{2}-\f12 m\om(t)^{2}q^{2}
\right]\,,
\ee
where the parameter $\om(t)$ a priori depends explicitly on time. The corresponding equation of motion is the standard second order differential equation describing an oscillator with a time-dependent frequency:
\be
\ddot{q}+\om(t)^{2}q=0
\,.
\ee
When the function $\om(t)$ is periodic, this is a Floquet oscillator. But, here, we do not make any assumption on the time dependence of the potential.

A method to integrate the motion is to perform a change of time variable, in order to simplify the equation of motion and remove the time dependence of the potential. More precisely, we would like to perform a time reparametrization allowing to map the equation of motion onto a standard harmonic oscillator with fixed constant frequency. This approach also puts in light the conformal symmetry of the mechanical system whatever the potential function $\om(t)$ one considers.
Let us introduce the following time reparametrization transformations:
\be
\label{eq:timereparam}
\left|\begin{array}{rcrcl}
t& \mapsto & \tt&=&f(t)\,, \\
q(t)& \mapsto & \tq(\tt)&=&h(t)^{\f12}q(t) \,,
\end{array}
\right.
\ee
with the Jacobian $h=\dot{f}$. These are diffeomorphisms of the time direction, for which the particle position is assumed to carry a $\f12$ conformal scaling dimension.
A quick calculation shows that that the action's shape is not modified by such a time reparametrization, up to a boundary term,
\be
s_{\tom}[\tt,\tq(\tt)]
=
s_{\om}[t,q(t)]
+
\int \rd t\,
\rd_{t}\left[
\f m4q^{2}\rd_{t}{\ln h}
\right]
\,,
\ee
though the potential function $\om(t)$ acquires a non-trivial transformation law,
\be
q=h^{-\f12}\,\tq\circ f
\quad\textrm{and}\quad
\om^{2}
=
h^{2}(\tom\circ f)^{2}+\f12\textrm{Schw}[f]
\,,
\ee
in terms of the Schwarzian derivative of the reparametrization function,
\be
\textrm{Schw}[f]=\rd_{t}^{2}{\ln h}-\f12(\rd_{t}{\ln h})^{2}
\,.
\ee
We recognize the well-known Virasoro group action on Sturm-Loiuville operators, see e.g. \cite{Ovsienko}.
A direct consequence is that solutions of the original equation of motion with $\om(t)$ give solutions of the mapped equation of motion with $\tom(\tt)$:
\be
\rd_{\tt}^{2}\tq+\tom(\tt)^{2}\tq=0
\quad\Leftrightarrow\quad
\rd_{t}^{2}q+\om(t)^{2}q=0
\,.
\ee

These time reparametrizations actually are symmetries of the system when they preserve the potential,
\be
f\textrm{ symmetry}\,\Leftrightarrow\,
\om=\tom\,.
\ee
If we consider the free system, i.e. with vanishing potential $\om=0$, symmetries are thus given by time reparameterizations with vanishing Schwarzian derivative. These correspond to Mobi\"us transformations of the time coordintate,
\be
\textrm{Schw}[f]=0
\,\Leftrightarrow\,
f(\tau)=\f{a \tau +b}{c\tau+d}
\,,
\ee
which form a $\SL(2,\R)$ Lie group. This defines the conformal symmetry under $\SL(2,R)$ transformations of the free system.
Now considering an arbitrary initial function $\om(t)$, assuming that we can identify a time reparameterization $F$ which maps the time dependent coupling $\om(t)$ to the free system $\tom=0$, then the system, with time-dependent harmonic potential $\f12\om(t) q^2$, also admits a $\SL(2,\R)$ symmetry group of transformations $F^{-1}\circ f_{M}\circ F$, where $f_{M}$ denotes the Mobi\"us transformations.
So, out of the whole $\mathrm{Diff}({\cal S}_{1})$ group of Virasoro time reparametrizations, only a $SL(2,\R)$  subgroup are actual symmetries of the systems. The remaining transformations are non-trivial maps between systems with different time dependent potentials.

\section{Synchronizing clocks}

In order to solve the system, let us use these time reparametrization to map an arbitrary time-dependent frequency $\om(t)$  to a constant frequency $\Om^{2}$. The frequency mapping equation becomes a Schwarzian equation for the reparametrization function:
\be
\om^{2}
=
h^{2}\Om^{2}+\f12\textrm{Schw}[f]
=
h^{2}\Om^{2}+\f{\ddot{h}}{2h}+\f{3\dot{h}^{2}}{4h^{2}}
\,.
\label{eqn:synchro}
\ee 
This is a non-linear second order differential equation on the Jacobian $h$ of the reparametrization map. It can be re-written in a simpler form by defining $\eta\equiv h^{-\f12}$, which leads to:
\be
\ddot{\eta}+\om^{2}\eta=\f{\Om^{2}}{\eta^{3}}\,.
\ee
This auxiliary differential equation is the original oscillating differential equation plus a non-linear source. Its solution defines a {\it synchronizing clock}, $\tau(t)=\int^t \rd t/\eta^2$, for the system is simply a harmonic oscillator with the constant frequency $\Om$.
{
More precisely, assuming that having identified a function $\eta$ satisfying the differential equation above, one can directly check that the time reparametrization map
\be
t\mapsto \tau(t)\,,\quad
q(t)\mapsto Q(\tau)=\eta^{-1}(t)q(t)\,,
\ee
is such that the initial equation of motion $\rd_{t}^{2}{q}+\om(t)^{2}q=0$ is equivalent to the harmonic oscillator evolution equation $\rd_{\tau}^{2}Q+\Om^{2}Q=0$ with constant $\Om$.
Let us point out that, since $\eta^{2}>0$ is strictly positive, the map $\tau(t)$ is by construction monotone and, thus, bijective.
}

We refer to such a clock as ``synchronizing'' since they map the irregular oscillations of the system driven by a time dependent potential $\om(t)$ to the regular oscillations of a fixed frequency harmonic oscillator.
%
%
We would like to stress that this is an important feature of time-dependent harmonic potential: being able to absorb all the time dependence of the potential and mapping the system onto a mere regular oscillator is an a priori surprising property. It is possible only due to the Schwarzian (cocycle) term of the Virasoro group action.

%
%
%

This construction, with synchronizing clocks, allows to define constants of motion for systems evolving in time dependent harmonic potentials. These are the Ermakov-Lewis invariants \cite{Lewis,Leach,LewisLeach,Schuch:2008tha,Gallegos2018,Padmanabhan2018}, studied in the next section.

\section{Ermakov invariants}

As soon as we have a solution $\eta$ to the auxiliary differential equation, then one can define the Ermakov-Lewis invariant \cite{Lewis1967}:
\be
2I_{\eta}[q]=
(\eta\dot{q}-\dot{\eta}q)^{2}+\f{\Om^{2}q^{2}}{\eta^{2}}
\,.
\ee
It is a constant of motion for the particle evolving in the time-dependent harmonic potential $V=\f12\om^{2}x^{2}$,
\be
\ddot{q}+\om^2q=0
\quad\Rightarrow\quad
\rd_{t}I_{\eta}[q]=0
\,.
\ee
As pointed out by Padmanabhan \cite{Padmanabhan2018,Gallegos2018}, it can be identified as the energy of the constant frequency harmonic oscillator in the synchronizing clock defined by the solution $\eta$. Indeed, let us perform the time reparametrization to the constant frequency, $(t,q(t),\om(t))\,\mapsto\,(\tau,Q(\tau),\Om)$:
\be
\tau=f(t)\,\quad \rd\tau=h(t)\rd t=\f{\rd t}{\eta(t)^{2}}\,\quad
Q(\tau)=\f{q(t)}{\eta(t)}\,
\ee
then one easily computes
\be
\rd_{\tau}Q=\eta^{2}\rd_{t}(\eta^{-1}q)=\eta\dot{q}-\dot{\eta}q\,,
\ee
where one recognizes the Wronskian of the pair $(q,\eta)$, or equivalently the Lie derivative of the two vector fields $q$ and $\eta$, and get:
\be
2I_{\eta}[q]=(\rd_{\tau}Q)^{2}+\Om^{2}Q^{2}
\,.
\ee
Here we uncover another layer. Underlining the deep and systematic relation between symmetry and constants of motion, we show that the Ermakov-Lewis invariants are actually the Noether conserved charges for the $\SL(2,\R)$ conformal symmetry of 1d particles in a harmonic potential.

Taking a step back, and calling $q_{1}$ and $q_{2}$ two independent solutions to the equation of motion $\ddot{q}+\om^2q=0$, it is well-known that the Wronskians with $q_{1}$ and $q_{2}$ provide two constants of motion,
\be
W_{a=1,2}[q]=q_{a}\dot{q}-\dot{q}_{a}q\,,\quad
\ddot{q}+\om(t)^{2}q=0\Rightarrow
\dot{W}_{a}=0\,.
\ee
Considering the canonical phase space structure $\{q,\dot{q}\}=\f1m$, the infinitesimal variation generated by the Wronskian is given the Poisson bracket $\delta_{a}q=\{q,W_{a}\}=q_{a}/m$. So the two Wronskians generate the classical symmetry under translations, $q\mapsto q+\eps_{a} q_{a}$, which simply reflects the linearity of the equation of motion.

It turns out that the Ermakov-Lewis invariants are linear combinations of the squared Wronskians.
The key point is that solving the auxiliary differential equation is entirely equivalent to solving the original differential equation with a time-dependent frequency. Indeed, the solutions to $\ddot{\eta}+\om^{2}\eta={\Om^{2}}{\eta^{-3}}$ are simply:
\be
\eta=(aq_{1}^{2}+bq_{2}^{2}+2cq_{1}q_{2})^{\f12}\,,
\ee
with the constant frequency $\Om^{2}=(ab-c ^{2})W^{2}$ given in terms of the Wronskian $W$ between the two solutions $q_{1}$ and $q_{2}$:
\be
W=q_{1}\dot{q}_{2}-\dot{q}_{1}q_{2}
\,,\quad
\rd_{t}W=0
\,.
\ee
Then a straightforward calculation yields the  result:
\be
2I_{\eta}=aW_{1}^{2}+bW_{2}^{2}+cW_{1}W_{2}\,.
\ee
As a quadratic polynomial in the Wronskians, $I_{\eta}$ is thus automatically a conserved charge.
We can compute the infinitesimal symmetry transformations generated by the three squared Wronskians,
\begin{align}
&\delta_{11}q=\{q,W_{1}^{2}\}=
\f2m(q_{1}^{2}\dot{q}-q_{1}\dot{q}_{1}q)\,,
\\
&\delta_{22}q=\{q,W_{2}^{2}\}=
\f2m(q_{2}^{2}\dot{q}-q_{2}\dot{q}_{2}q)\,,
\\
&\delta_{12}q=\{q,W_{1}W_{2}\}=
\f1m(2q_{1}q_{2}\dot{q}-\dot{q}_{1}q_{2}q-q_{1}\dot{q}_{2}q)\,.
\end{align}
These are special cases of more general transformations generated by arbitrary vector fields $X=X(t)\pp_{t}$,
\be
\delta_{X} q=X\dot{q}-\f12\dot{X}q
\,,
\ee
for $X$ equal respectively to $q_{1}^{2}$, $q_{2}^{2}$ and $q_{1}q_{2}$.
The variations $\delta_{X}$ correspond to the infinitesimal time reparameterizations  $t\mapsto f(t)=t+\eps X(t)$,
 with the $\f12$ factor corresponding to the conformal weight of the position coordinate $q$ in the transformations \eqref{eq:timereparam}.
Indeed, making sure not to forget the Jacobian factor $\dot{f}=1+\eps \dot{X}$, we get:
\beq
\tq(t)-q(t)
&=&
-\eps\left(
X\dot{q}-\f12\dot{X}q
\right)
\,.
\eeq
Applying such infinitesimal transformations to the Lagrangian $L_{\om}=\f m2(\dot{q}^{2}-\om^2q^{2})$, we get the following variation:
\beq
&&\delta_{X}L_{\om}
=
\rd_{t}
\cK
+\f m4q^{2}\Big{(}X^{(3)}+4\om^{2}\dot{X}+4\om\dot{\om}X\Big{)}\,,
\nn\\
&&\textrm{with}\quad
\cK
=
\f m2
\left(
X\dot{q}^{2}-\f12\ddot{X}q^{2}
-\om^{2}Xq^{2}
\right)
\,.
\eeq
Vector fields $X$ generating symmetries of the system thus correspond to solutions of the  linear differential equation $X^{(3)}+4\om^{2}\dot{X}+4\om\dot{\om}X=0$. One easily checks that the three solutions to this third order differential equation are $X=q_{1}^{2}$ or $q_{2}^{2}$ or $q_{1}q_{2}$.
This confirms that the Ermakov-Lewis invariants, given by the squared Wronskians, are indeed Noether charges for the time reparametrization symmetry of the system.

Moreover, one can compute the Poisson bracket between those conserved charges. We get a canonical bracket between the two Wronskians,
\be
\{W_{1},W_{2}\}=\f1m W\,,
\ee
where we remind the reader that $W=(q_{1}\dot{q}_{2}-\dot{q}_{1}q_{2})$ is a constant, $\rd_{t}W=0$. Thus the squared Wronskians form a $\sl(2,\R)$ Lie algebra,
\begin{align}
&\{W_{1}W_{2},W_{1}^{2}\}=-\lambda W_{1}^{2}
\,,\quad
\{W_{1}W_{2},W_{2}^{2}\}=+\lambda W_{2}^{2}
\,,\nn
\\
&\{W_{1}^{2},W_{2}^{2}\}=2\lambda W_{1}W_{2}
\,,\quad
\textrm{with} \quad\lambda=2W/m
\,.
\end{align}
This confirms that, whatever the value of the time-dependent coupling $\om(t)$, the subgroup of time reparametrizations which are symmetries of the system  indeed form a $\SL(2,\R)$ group.
In fact, this is a universal mechanism  for 2nd order linear differential equations. For instance, it leads to the  invariance under $\SL(2,\R)$ conformal transformations of the FLRW cosmology background dynamics \cite{BenAchour:2019ufa,BenAchour:2020xif} and of black holes' Love numbers \cite{BenAchour:2022uqo,Berens:2022ebl}.

%
 
%
%
%
 
\medskip

Now, the goal of this letter is to show that the existence of synchronizing clocks and the $\SL(2,\R)$ invariance, which are entirely classical questions, both naturally stem from quantum mechanics.
Indeed, it turns out that the evolution of the quantum uncertainty of the quantized system, defined by the standard Schr\"odinger equation, naturally provides solutions to this classical problem.

\section{Quantum uncertainty dynamics}

Let us quantize the system into  a 1d wave-function $\psi(t,x)$ driven by the Schr\"odinger equation with time-dependent potential:
\be
i\hbar \pp_{t}\psi=-\f{\hbar^{2}}{2m}\pp_{x}^{2}\psi+\f12m\om(t)^{2}x^{2}\psi
\,.
\ee
The wave-function $\psi(x)$ is a field and carries an infinite number of degrees of freedom, one per space point $x$, compared to the classical mechanical system with a single degree of freedom $q$.
In the semi-classical regime, for wave-functions peaked around their mean values in position and momentum\footnote{Semi-classical wave-functions are usually defined as coherent wave-packets. Then a general wave-function can be written as a superposition of such coherent states.}, the classical motion is recovered from the expectation values of the position and momentum operators. Then the infinite number of degrees of freedom, carried by the wave-function, dress the classical motion with extra degrees of freedom given by all the moments of the wave-functions. In the semi-classical regime, these moments describe the shape and fluctuation of wave-packets around the classical trajectory.
%

As a leading-order approximation, one can focus on the quadratic moments of the wave-function and put aside higher moments. This is modeled by considering a Gaussian ansatz for the wave-function. Such Gaussian ansatz depend on a global pre-factor (normalization and phase), the mean position, momentum, and (complex) spread. This reduces the infinite number of degrees of freedom of the wave-function back to a finite number of dynamical variables, so that this truncation of quantum mechanics can be written as an effective classical mechanical system, where the classical position and momentum now evolve together with the position and momentum spread of the wave-packet. It turns out that this truncation is actually exact for a harmonic potential, even if time-dependent. Indeed, in that case, Gaussian states are exact solutions of the Schr\"odinger equations, which translate into equations of motion for the Gaussian parameters.

More precisely, we  focus on the dynamics of a Gaussian wave-packet:
\be
\psi_{Gaussian}(t,x)=Ne^{i\gamma}e^{i\f{xp}{\hbar}}e^{-A(x-q)^{2}}\,,
\ee
where $\gamma(t)$ is a global phase, $p(t)$ is the momentum, $q(t)$ the average position and $A(t)\in\C$ controls the uncertainty. 
We parametrize the complex Gaussian width as in \cite{Gaussianwavepacket}:
\be
A=\f1{4\alpha^{2}}\left(
1-i\f{2\alpha\beta}{\hbar}
\right)
\,.
\ee
The normalization factor $N>0$ is fixed by requiring that the wave-packet be normalized:
\be
1=\int \rd x\,|\psi|^{2}=N^{2}\alpha\sqrt{2\pi}\,.
\ee
The physical interpretation of the parameters $q$,$p$ and $A$ is validated by computing the expectation values of the linear and quadratic combinations of the position operator $\hat{x}$ and the momentum operator $\hat{p}=-i\hbar\pp_{x}$ :
\beq
&&\la\hat{x}\ra=q
\,,\quad
\la\hat{x}^{2}\ra
=
q^{2}+\alpha^{2}
\,,
\\
&&\la\hat{p}\ra=p
\,,\quad
\la\hat{p}^{2}\ra=
p^{2}+\beta^{2}+\f{\hbar^{2}}{4\alpha^{2}}
\,,
\nn\\
&&\hat{D}\equiv \f12(\hat{x}\hat{p}+\hat{p}\hat{x})
\,,\quad
\la \hat{D}\ra=pq+\alpha\beta
\,.\nn
\eeq
Plugging the Gaussian ansatz in the Schr\"odinger equation shows that it is an exact  solution if and only if the following equations of motion holds:
\beq
&&\dot{q}=\f pm
\,,\qquad
\dot{p}=-m\om^{2}q\,,
\\
&&\dot{\alpha}=\f\beta m
\,,\qquad
\dot{\beta}=
-m\om^{2}\alpha+\f{\hbar^{2}}{4m\alpha^{3}}\,,
\nn\\
&&\dot{\gamma}=-\f{\cL}\hbar-\f\hbar{4m\alpha^{2}}\,,\quad
\textrm{with}\,\,
\cL=\f{p^{2}}{2m}-\f12m\om^{2}q^{2}\,.\nn
\eeq
This means that the position uncertainty $\alpha$ satisfies exactly the auxiliary non-linear differential equation for a synchronizing clock:
\be
\ddot{\alpha}
+\om^{2}\alpha=\f{\Om^2}{\alpha^{3}}
\quad\textrm{with}\,\,
\Om=\f{\hbar}{2m}
\,.
\ee
The corresponding synchronizing time coordinate is thus the integral of the real part of the Gaussian width of the wave-packet:
\be
\tau
=
\int^{t}\rd t\,\f1{\alpha^{2}}
=
4\int^{t}\rd t\,\textrm{Re}(A)
\,.
\ee
So the quantum uncertainty
can be used as an intrinsic clock for the system, which makes it tick regularly as an harmonic oscillator at fixed frequency. This is the main result of the present letter.

\bigskip

Moreover, we would like to remind the reader that the quantum uncertainty is a variable on its own in quantum mechanics, on top of the expectation values of position and momentum. The quadratic spread in position and momentum reflect the shape of the wave-packets around the expectation values. Those spread evolve, fluctuate and can be excited. These are the squeezing modes.

Here, we see from the effective equations of motion above that the quantum dynamics of a Gaussian wave-packet is equivalent to a doubled classical phase space,
\be
\{q,p\}=\{\alpha,\beta\}=1\,,
\ee
provided with an effective dynamics taking into account the dynamics of the quantum uncertainty \cite{Gaussianwavepacket,Prezhdo2006QuantizedHD,Baytas:2018gbu,Bojowald:2022lbe}:
\be
H_{eff}[p,q,\alpha,\beta]
=
H[p,q]+H_{eff}^{(2)}[\alpha,\beta]
\,,
\ee
with the classical Hamiltonian $H[p,q]
=
\f{p^{2}}{2m}+\f m2\om^{2}q^{2}$ and 
\be
H_{eff}^{(2)}
=
\f{\beta^{2}}{2m}+\f12m\om^{2}\alpha^{2}+\f{\hbar^{2}}{8m\alpha^{2}}
\,.
\ee
Although the extra Poisson pair seems to pop out of nowhere, the canonical bracket $\{\alpha,\beta\}=1$ actually descends from the field Poisson bracket for Schr\"odinger's field theory, when writing 1-dimensional quantum mechanics as a 1+1-dimensional non-relativistic field theory whose field equation is simply Schr\"odinger equations. We will briefly overview this formalism in the next section. The interested reader can find a more in-depth analysis in \cite{Livine:2023vph,Livine:2023qvx}.

In some sense, focussing on the Gaussian wave-packet ansatz defines a ``Gaussian mini-superspace'' for quantum mechanics, similarly to what's usually done for cosmological and astrophysical applications of general relativity, e.g. \cite{Vilenkin:1994rn}.
The logic here, and developed earlier in \cite{Gaussianwavepacket,Bojowald:2005cw,Prezhdo2006QuantizedHD}, is similar to effective field theories extracting order by order the corrections due to quantum or gravitational effects, for example in a post-Newtonian perturbative scheme for quantum mechanics in a given gravitational background \cite{PhysRevD.44.1067,LAMMERZAHL199512,Buoninfante:2021qrv}. Let us nevertheless stress that the effective Hamiltonian derived here comes directly from the Schr\"odinger action for the wave-function with no gravity or other input.

We have clearly identified the Gaussian width parameters $(\alpha,\beta)$ as extra parameters of the wave-packet, beside the average position and momentum $(q,p)$. We have written their equations of motion, descending from Schr\"odinger equation, and reminded the reader that they evolve according to an effective Hamiltonian. This supports the semi-classical interpretation of a quantum particle as a dressed particle, carrying extra degrees of freedom on top of the classical position and momentum. These are the higher moments of the wave-functions and encode the shape of the wave-packet around the linear expectation values. The quadratic moments, or quadratic quantum uncertainty, as studied here, describe the squeezing of the wave-packet and follow effective classical equations of motion. For a classical harmonic potential, they are shown to almost follow the same dynamics as the classical system, but for an extra potential term in $\alpha^{-2}$ in the effective Hamiltonian.
It is interesting to note that this is not a $\alpha^{-1}$ potential as for a gravitational or electric force, but a $\alpha^{-2}$ potential as for the angular momentum term for 3d particles. It is usually referred to as a conformal potential, e.g. \cite{deAlfaro:1976vlx}. This new potential term grows as the wave-packet is more peaked in position; it is responsible for the diffusive behavior of quantum mechanics, which forces  wave-packets to spread out unless another force counters it and localizes the wave-packets. Here that counter-diffusive force is the harmonic potential. From this perspective, it is this extra effective conformal potential that
is responsible for ensuring that the Heisenberg uncertainty principle holds, $(\delta x)(\delta p)\ge \hbar/2$.

This logic can be extended to the higher moments of the wave-function: the Schr\"odinger equation endows them with a classical evolution and they can also be used as internal clocks for the system. This is studied in a technical companion paper \cite{Livine:2023vph}. Let us point out that higher order potentials, e.g. with a quartic term in $q^4$, generate couplings between the various moments, in particular they couple the evolution of the position and momentum expectation values to the quadratic uncertainty \cite{Gaussianwavepacket,Bojowald:2005cw,Prezhdo2006QuantizedHD}.

\section{Ermakov Invariant as  Uncertainty}

A vantage point to better understand the mechanisms at hand is the field theory point of view. Indeed, considering the wave-function $\psi$ as a classical field, the Schr\"odinger equation becomes the equation of motion of a classical field theory. This is Schr\"odinger field theory (see e.g. \cite{Padmanabhan:2017bll}), whose action is:
\beq
&\cS[\psi]
=
\int \rd t\left[
\int \rd x\,i\hbar \bpsi\pp_{t}{\psi}-\cH[\psi]
\right],&
\\
&\textrm{with}\quad
\displaystyle{\cH[\psi]
=
\int \rd x\left[
\f{\hbar^{2}}{2m}\pp_{x}\bpsi\pp_{x}\psi+V(t)|\psi|^{2}
\right],}&\nn
\eeq
with the time-dependent potential $V(t)=\f m2\om(t)^{2}x^{2}$.
The theory is directly written in Hamiltonian form. The field Poisson bracket is $i\hbar\{\psi(x),\bpsi(y)\}=\delta(x-y)$ and the evolution of observables is given by the Poisson bracket with the  field Hamiltonian $\cH$.

Inserting the Gaussian wave-packet ansatz in this action directly leading to the doubled phase space and effective Hamiltonian derived earlier:
\beq
\cS[\psi]
&=&
\int
\rd t\,\bigg{[}
p\dot{q}+\beta\dot{\alpha}
-H_{eff}[p,q,\alpha,\beta]
\bigg{]},
\eeq
up to a total derivative $\rd_{t}\left(\f12 \alpha\beta -\hbar \gamma\right)$.
The phase $\gamma$ does not have a conjugate momentum because we fixed the number of particles by normalizing the wave-function, $\int \rd x\,|\psi|^{2}=1$.

The expectation values of the quadratic polynomials of the position and momentum operators are field observables, 
\beq
&&\la \hat{x}^{2}\ra
=\int\rd x\,x^{2}\bpsi\psi
\,,\quad
\la \hat{p}^{2}\ra
=-\hbar^{2}\int\rd x\,\bpsi\pp_{x}^{2}\psi
\,,\nn\\
&&\displaystyle{\la \hat{D}\ra
=
-i\hbar\int\rd x\,\bpsi(x\pp_{x}+\pp_{x}x)\psi\,.}
\eeq
As well-known, they  form a $\sl_{2}(\R)$ algebra, which we refer to as the  {\it uncertainty algebra},
\be
\begin{array}{lcl}
\{\la \hat{x}^{2}\ra,\la \hat{p}^{2}\ra\}
&=&
4\la \hat{D}\ra
\,,\\
\{\la \hat{D}\ra,\la \hat{x}^{2}\ra\}
&=&
-2\la \hat{x}^{2}\ra
\,,\\
\{\la \hat{D}\ra,\la \hat{p}^{2}\ra\}
&=&
2\la \hat{p}^{2}\ra
\,.
\end{array}
\ee
This structure allows us to generalize the $(\alpha,\beta)$ beyond the Gaussian truncation. Indeed us define for arbitrary wave-functions:
\be
\alpha\equiv \sqrt{\la \hat{x}^{2}\ra-\la \hat{x}\ra^{2}}
\,,\quad
\alpha\beta\equiv\la \hat{D}\ra-\la \hat{x}\ra\la \hat{p}\ra
\,.
\ee
A quick computation using the $\sl(2,\R)$ brackets above, or directly the field Poisson bracket, gives:
\be
\{\alpha,\beta\}=1
\,,
\ee
thereby showing that the ``extra'' canonical pair $(\alpha,\beta)$ is not an artefact of the Gaussian ansatz but reflects the degrees of freedom encoded in the moments of the wave-function.

The $\sl(2,\R)$ algebraic structure of the quadratic moments reflects the fact that their equations of motion form a closed set of differential equations,
\be
\rd_{t}\la \hat{x}^{2}\ra=\f{2\hbar^{2}}{m}\la \hat{D}\ra\,,
\quad
\rd_{t}\la \hat{p}^{2}\ra=-2\hbar^{2}m\om^{2}\la \hat{D}\ra\,,
\ee
\be
\rd_{t}\la \hat{D}\ra
=
\f{\hbar^{2}}m\rd_{t}\la \hat{p}^{2}\ra
-\hbar^{2}m\om^{2}\la \hat{x}^{2}\ra
\,.
\ee
A direct consequence is the following constant of motion,
\be
C\equiv
\la \hat{x}^{2}\ra\la \hat{p}^{2}\ra-\la \hat{D}\ra^{2}\,,\qquad
\rd_{t}C=0
\,.
\ee
This observable encodes the quantum uncertainty of the wave-function at quadratic order.
It is in fact the Casimir of the $\sl_{2}(\R)$ algebra, so one can call it the {\it uncertainty Casimir}.  It is indeed a constant of motion,  since the Hamiltonian $\cH[\psi]$ belongs to the $\sl_{2}(\R)$ algebra,
\be
\cH[\psi]
=
\f1{2m}\la \hat{p}^{2}\ra+\f m2 \om^{2}\la \hat{x}^{2}\ra
\,.
\ee
This uncertainty Casimir reflects the uncertainty in both position and momentum. Evaluating it on the Gaussian wave-packet gives:
\be
C=(\alpha p -\beta q)^{2}+\f{\hbar^{2}}{4\alpha^{2}}q^{2}+\f{\hbar^{2}}{4}
\,.
\ee
which matches exactly with the Ermakov-Lewis invariant for $\eta=\alpha$:
\be
I_{\alpha}
=
\f1{2m^{2}}\left[
(\alpha p -\beta q)^{2}+\f{\hbar^{2}q^{2}}{4\alpha^{2}}
\right]\,.
\ee
This shows that the Ermakov-Lewis invariant, which is the classical Noether charge for the conformal invariance of a time-dependent classical oscillator, is directly equal to the quantum uncertainty for the quantized oscillator.
The fact that it is a constant of motion directly follows from the coupled evolution of the position and momentum expectation values and their quadratic spread, all driven by the same Schr\"odinger equation dictating the dynamics of the whole wave-function.
Having a quantum observable playing the role of a classical symmetry generator is an unexpected equivalence between classical and quantum notions, which echoes the main result that the quantum uncertainty plays the role of a synchronizing clock for the classical system.

\section*{Conclusion \& Outlook}

To summarize, we have mapped the evolution of a system in a time-dependent harmonic potential onto a standard regular harmonic oscillator by looking at its dynamics with respect to the position spread of the wave packet of the quantized system. 
This shows that using the quantum uncertainty as a clock can be a useful mathematical technique from quantum mechanics to solve a classical mechanics problem.
We believe it is actually a deeper concept.
This underlines that the shape of the wave-function is not a static dressing of the classical motion but is instead a truly dynamical objects which evolves and fluctuates independently.
It suggests a new take on the physical role of quantum uncertainty as an intrinsic clock.
It opens the door to studying the (evolution of the) correlations between the wave-function main expectation values and their higher moments, or, in looser terms, between the  classical degrees of freedom and their quantum dressing. And it underlines the  possibility of constructing relational observables between systems and their own quantum fluctuations, in quantum field theory, quantum information and quantum gravity.

At a more technical level, the presented analysis has clarified the role of the Ermakov-Lewis invariant for oscillators in a  time-dependent harmonic potential. On the classical side, we have shown that those invariants generate the conformal symmetry of classical mechanics under $\sl_{2}(\R)$ transformations. Then we have highlighted the quantum origin of those invariants, by proving that they match the quantum uncertainty measured from the quadratic moments of the wave-functions, $\la {x}^{2}\ra\la {p}^{2}\ra-\la xp\ra^{2}$, for Gaussian wave-packets evolving in the time-dependent harmonic potential.

Beyond technical improvements such as the necessary extension to the analysis of the dynamics of higher multipoles of the wave-packets (representing the non-Gaussianities) (see the companion work \cite{Livine:2023vph}), we see two main arenas of application.
On the one hand, it would be great to design a quantum information experiment to check the relative evolution of the position and position uncertainty for a time dependent harmonic potential, if one has access to an efficient non-destructive measurement of the quadratic uncertainties of a wave-packet (the squeezing modes). This would allow for an experimental check that the wave-packet spread indeed provides a synchronizing clock for the classical dynamics in the time-dependent harmonic well. It would also allow to explore further the unveiled relational evolution of the classical position with respect to the quantum uncertainty and the resulting possible existence of a universal self-synchronization process.

On the other hand, it would be enlightening to push the present logic further and develop its applications to classical and quantum gravity (see \cite{deBoer:2022zka,Draper:2022pvk,Faulkner:2022mlp,Carney:2022dku,Zurek:2022xzl,Harlow:2022qsq} for recent reports on the theoretical and experimental progress on quantum gravity, realized in the framework of the Snowmass 2021 study), for instance in the context of holography and conserved charges where soft mode dressing plays a very similar role to the wave-function higher moments \cite{Giddings:2019hjc}.
An enticing line of research would be to tackle the problem of time\footnote{One of the main issue in quantum gravity is the problem of localization and, in particular, the problem of time \cite{Isham:1992ms}, which is especially relevant to quantum cosmology.
This directly stems from the symmetry of general relativity under space-time diffeomorphisms. This gauge invariance, which is the mathematical expression of the relativity principle stating that the laws of physics should be the same for all observers, is both the paradigm-changing insight of the theory and the central obstacle to defining physical localized observables, e.g. \cite{Giddings:2005id,armas_2021}.
It is both a conceptual and practical problem.
Standard approaches are bulk reconstruction from  boundary charges and  boundary conformal field theory (in particular in the AdS/CFT framework), see e.g. \cite{Czech:2016xec,Giddings:2019hjc}, and the introduction of relational observables \cite{Tambornino:2011vg}, which encode the evolution of a subsystem in terms of another subsystem and led to a renewed interest in quantum reference frame from a quantum gravity perspective \cite{Carrozza:2022xut,Goeller:2022rsx}.
} in quantum gravity, exploiting the intriguing possibility of a notion of intrinsic clock  due to quantum degrees of freedom: we would look  at the classical evolution of a system in terms of a clock built built from its own quantum fluctuations. It suggests that going fully quantum in quantum gravity, instead of focussing on building classical relational observables, might be an interesting route to discuss the problem of time. In layman terms, we would look at the evolution of the cosmological long wave length modes in terms of the dynamics and ticking of the Planckian modes.
More specifically in cosmology, matter fields evolve at leading order as in a time-dependent harmonic potential whose time dependence comes from the expansion of the universe. The perspective developed here suggests to   look at the evolution of the cosmological homogeneous background in terms of the local clock defined by the quantum fluctuations of the inhomogeneities, instead of the usual point of view of looking at the evolution of the inhomogeneities in terms of a global time coordinate defined by the background geometry.
%
The challenge would now be to turn this into a concrete ad practical proposal.
%

\section*{Acknowledgement}
I am  grateful to  Jibril Ben Achour and Carlo Rovelli for  discussions on symmetries, field theory and quantum mechanics.

This research was supported in part by Perimeter Institute for Theoretical Physics. Research at Perimeter Institute is supported by the Government of Canada through the Department of Innovation, Science and Economic Development and by the Province of Ontario through the Ministry of Research, Innovation and Science.




\bibliographystyle{bib-style}
\bibliography{QM}

\end{document}